\documentclass[final]{svjour2}
\usepackage{graphicx}
\usepackage{sidecap}
\usepackage{rotating}
\usepackage{amssymb}
\usepackage{epstopdf} 
\usepackage{mathptmx}
\usepackage[numbers]{natbib}
\def\bra#1{\bigl\langle{ #1} \bigr|}
\def\ket#1{\bigl|{ #1} \bigr\rangle}

\def\rvec {{\bf r}}

\def\kvec {{\bf k}}

\def\he#1{$^{#1}$He}
\def\EF{e_{\rm F}}
\def\KF{k_{\rm F}}
\def\SF{S_{\rm F}}
\def\a0{a_0}
\def\I{{\rm i}}

\usepackage{soul, xcolor} 
\setstcolor{red}

\makeatletter
\journalname{Journal of Low Temperature Physics}

\bibpunct{}{}{,}{s}{}{,}

\begin{document}

\newcommand{\hdblarrow}{H\makebox[0.9ex][l]{$\downdownarrows$}-}

\title{Pairing of the P\"oschl-Teller gas}

\author{H.-H. Fan$^{\dagger\ddagger}$ \and E. Krotscheck$^{\dagger\ddagger}$}
\institute{$^\dagger$Department of Physics, University at Buffalo SUNY,
Buffalo, NY 14260, USA\\
  $^\ddagger$Institut f\"ur Theoretische Physik, Johannes
Kepler Universit\"at, A 4040 Linz, Austria\\
}

\maketitle

\begin{abstract}
  
  We report calculations of equation of state of a model system,
  representative for a cold Fermi gas, of particles interacting via
  the P\"oschl-Teller interaction. In successively more sophisticated
  calculations, we determine the importance of correlations and
  non-localities. We show that the gas displays, at relatively low
  density, an instability indicated by a divergence of the {\em
    in-medium\/} scattering length which occurs well before the
  divergence of the vacuum scattering length and the spinodal density.
  We also calculate the pairing gap and show that non-local
  correlations can change the pairing gap by almost an order of
  magnitude.
  
  \keywords{Superfluidity, Quantum Fluids, Cold Gases}

\end{abstract}

\section{Motivation} 
\label{sec:intro}

In the analysis of experiments with cold quantum (Bose- and
Fermi-)gases, one is normally interested in a density regime where the
short-ranged details of the interaction are irrelevant, and the
structure and dynamics of the system can be characterized by
low-energy 2-body scattering processes. Much of the physics -- with
rare exceptions like the ``unitary limit'' -- in this regime can be
described by textbook methods, and one of the tasks of microscopic
many-body theory is to explore the regime of validity of such
approaches. To that end, we examine in this work the equation of
state, the stability, and the pairing properties of a low-density
Fermi gas interacting via a purely attractive, short-ranged
interaction.

\section{Generic Many-Body Theory}
\label{sec:GMBT}

Our method of choice is the Jastrow-Feenberg \cite{FeenbergBook}
variational approach; pedagogical and review material may be found in
Refs. 
\citenum{FeenbergBook}-\nocite{Johnreview}\citenum{KroTriesteBook};
technical details are in Ref. \citenum{polish}.

We assume for a strongly interacting and translationally invariant {\em
  normal\/} system a non-relativistic many-body
Hamiltonian of identical particles interacting via a pair-potential
$V(\rvec_i-\rvec_j)$.  The method starts with an {\em ansatz\/} for
the wave function
\begin{eqnarray}
\Psi_0({\bf r}_1,\ldots,{\bf r}_N) &=& I^{-1/2}_{{\bf o},{\bf o}}
	F({\bf r}_1,\ldots,{\bf r}_N)
	\Phi_0(1,\ldots,N)\label{eq:wavefunction},\\
F({\bf r}_1,\ldots,{\bf r}_N) &=& \exp\frac{1}{2}
\left[\sum_{i<j}  u_2({\bf r}_i,{\bf r}_j) + \cdot\cdot +  
\sum_{i_1<\ldots<i_n}u_n({\bf r}_{i_1},.., {\bf r}_{i_n}) 
+ \cdot\cdot \right]\,,
\label{eq:Jastrow}
\end{eqnarray}
where ${I_{{\bf o},{\bf o}}} = \left\langle \Phi_0 | F^\dagger F |
\Phi_0\right\rangle$ is the normalization constant.
$\Phi_0(1,\ldots,N)$ is a model state, which is for normal Fermi
systems a Slater-determinant, and $F$ is a local correlation operator
written in the general form (\ref{eq:Jastrow}).  Diagrammatic methods,
specifically the optimized Euler-Lagrange Fermi-hypernetted chain
(FHNC-EL) method, determine the correlations $u_n({\bf
  r}_1,\ldots,{\bf r}_n)$ by functionally minimizing the energy.
Truncated at the two-body term $u_2$, it is the standard Jastrow
approximation. These are computationally far less demanding than
quantum Monte Carlo calculations and, more importantly, provide direct
information on physical processes. For example phase transitions are
reflected in divergences of the theory. The resulting equations are
equivalent to the summation of localized parquet--diagrams
\cite{parquet1} and can also be derived from Coupled Cluster theory
\cite{BishopValencia} without ever mentioning a Jastrow function. For
that reason, we refer to our method as ``generic.''

For our specific application, we assume here the ``P\"oschl-Teller''
interaction \cite{PoeschlTeller}
\begin{equation}
  V(r) = - \frac{\hbar^2}{m\sigma^2}
  \frac{V_0(V_0-1)}{\cosh^2(r/\sigma)}
  \label{eq:Vpt}
\end{equation}
characterized by the strength $V_0$ and the range $\sigma$.  For this
interaction, the scattering length can be calculated analytically
\cite{Fluegge}. In the above parametrization, bound states appear at
even integer values of $V_0$; the regime of interest is therefore the
range $1< V_0 < 2$. This interaction has been used by Gezerlis and
Carlson \cite{GC2008} for a Monte Carlo study of strong pairing in
cold gases, it is similar to the attractive square--well potential
used by Astrakharchik {\em et al.}  \cite{astraPRL04} and us
\cite{cbcs}.

We spell out the simplest versions of the equations that are
consistent with the variational problem (``FHNC//0-EL''). These
equations provide the {\em minimal\/} version of the FHNC-EL
theory. In particular, they contain the indispensable physics, namely
the correct description of both short- and long-ranged correlations.

The minimization condition for the pair correlations can be written 
in the form
\begin{equation}
S(k) = \SF(k)\left[1 +
	2\frac{\displaystyle \SF^2(k)}{\displaystyle t(k)}
\tilde V_{\rm p-h}(k)\right]^{-\frac{1}{2}}\,,
\label{eq:FermiRPA0}
\end{equation}
where $S(k)$ is the static structure factor of the interacting system,
$\SF(k)$ is the static structure factor of the non-interacting Fermi
system, and $t(k) = \hbar^2 k^2/2m$.  In the FHNC//0 approximation, we
have
\begin{eqnarray}
V_{\rm p-h}(r) &=&
\left[1+ \Gamma_{\!\rm dd}(r)\right]V(r)
+ \frac{\hbar^2}{m}\left|\nabla\sqrt{1+\Gamma_{\!\rm dd}(r)}\right|^2
+ \Gamma_{\!\rm dd}(r)w_{\rm I}(r)
\label{eq:VddFermi0}\,,\\
\widetilde w_{\rm I}(k) &=& -t(k)
\left[\frac{1}{\SF(k)}-\frac{1}{ S(k)}\right]^2
\left[\frac{S(k)}{\SF(k)}+\frac{1}{2}\right]
\label{eq:inducedFermi0}\\
{\widetilde \Gamma}_{\!\rm dd}(k) &=& \bigl(S(k)-\SF(k)\bigr)\SF^{-2}(k)\,.
\label{eq:GFHNC}
\end{eqnarray}
We define the Fourier transform with a density factor, {\em i.e.},
$\tilde f(\kvec) \equiv \rho \int d^3r\, e^{\I\kvec\cdot\rvec}
f(\rvec)$\,.

More complicated versions of the FHNC-EL method add additional
equations for the so-called ``ee'', ``de,'' and ``cc'' diagrams
\cite{polish}.

\subsection{Correlated Basis Functions}
\label{ssec:CBF}
Eq. (\ref{eq:FermiRPA0}) can be interpreted in terms of linear
response theory as follows:
Begin with the random
phase approximation for the static structure function
  \begin{equation}
    S(k) = - \Im m \int_0^\infty \frac{d\omega}{\pi}
    \chi_0(k,\omega)\left[1-\tilde V_{\rm
        p-h}(k)\chi_0(k,\omega)\right]^{-1}\label{eq:SRPA}
  \end{equation}
  where $\chi_0(k,\omega)$ is the Lindhard function, and $\tilde
  V_{\rm p-h}(k)$ is a local quasiparticle interaction or
  ``pseudopotential\cite{Aldrich}''. Eq. (\ref{eq:FermiRPA0}) can be
  obtained by approximating the Lindhard function $\chi_0(q,\omega)$
  by a ``collective'' Lindhard function which is constructed by
  approximating the particle-hole band by an single pole such that the
  $m_0$ and $m_1$ sum rules of the non--interacting system are
  satisfied \cite{Rip79,polish}. A way to go beyond local correlation
  operators of the form (\ref{eq:Jastrow}) is then to replace the
  ``collective'' expression (\ref{eq:FermiRPA0}) by Eq.
  (\ref{eq:SRPA}), keeping the same particle-hole interaction. With
  this we go {\em beyond\/} the Jastrow-Feenberg approximation.  A
  rigorous proof that this procedure is legitimate can be obtained
  within correlated basis functions (CBF) theory
  \cite{rings,KroTriesteBook}.
 
\subsection{BCS Theory with correlated wave functions}
\label{ssec:CBCS}

The natural generalization of the Jastrow-Feenberg approach to a
superfluid system is to first project the bare BCS state on a complete
set of independent-particle states with fixed particle number. Then
apply the correlation operator to that state, normalize the result,
and, finally, sum over all particle numbers $N$. Thus, the correlated
BCS (CBCS) state becomes
\begin{equation}
\ket{\rm CBCS} =  \sum_{{\bf m},N} \ket {\Psi_{\bf m}^{(N)}}
\langle\Phi_{\bf m}^{(N)} \ket{\rm BCS} \,,\quad
\left|{\rm BCS} \right\rangle =
{\prod_{\kvec}}
\left[ u_{\kvec} +  v_{\kvec} a_{ {\bf  k} \uparrow }^\dagger
 a_{-{\bf  k} \downarrow}^\dagger  \right] |0 \rangle
\label{eq:BCS}
\end{equation}
where the $\ket{\Psi_{\bf m}^{(N)}} \equiv I^{-1/2}_{{\bf m},{\bf m}}
F_N\ket{\Phi_{\bf m}^{(N)}}$ with $ I_{{\bf m},{\bf m}} =
\bra{\Phi_{\bf m}^{(N)}}F_N^\dagger F_N\ket{\Phi_{\bf m}^{(N)}}$ form
a complete set of normalized, but non-orthogonal $N$-particle basis
states built with the correlation operator $F$ and the Slater
determinant $\ket{\Phi_{\bf m}^{(N)}}$, and $u_\kvec$, $v_\kvec$ are
the familiar Bogoliubov amplitudes.

Considering the interaction of only one Cooper pair at a time, we can
expand all expectation values in the deviation $u_\kvec$, $v_\kvec$
from their normal state values. (In fact, the inclusion of superfluid
momentum distributions to all orders has a rather small effect on the
pairing gap \cite{FanThesis} even when the gap is comparable to the
Fermi energy.) Then, all ingredients of the theory can be calculated
for the normal system.

The calculation of expectation values for correlated states is
somewhat tedious\cite{KroTriesteBook,cbcs}; we only give the final
result in the approximation used in this work.  The energy of the
superfluid state becomes
\begin{eqnarray}
\langle \hat H - \mu \hat N \rangle_s &=& H_{{\bf o},{\bf o}}^{(N)} - \mu N 
+ 2 \sum_{\kvec ,\,|\, \kvec \,|\,>\KF} v_{\kvec}^2 (t(k) - \mu ) 
- 2 \sum_{\kvec , \,|\, \kvec \,|\,<\KF} u_{\kvec}^2 (t(k) - \mu ) \nonumber \\ 
&\quad & + \sum_{\kvec,\kvec'}u_\kvec v_\kvec u_{\kvec'} v_{\kvec'} 
{\cal P}_{\kvec\kvec'}
\label{eq:Ebcs}
\end{eqnarray}
in terms of the ``pairing interaction'' of the form
\begin{eqnarray}
  {\cal P}_{\kvec\kvec'} &=& \bra{\kvec \uparrow ,-\kvec\downarrow}
  {\cal W}(1,2)\ket{\kvec'\uparrow ,-\kvec'\downarrow}\nonumber\\
  &&+ (|t(k)- \mu | 
+ |t(k')- \mu |)\bra{\kvec \uparrow ,-\kvec\downarrow}
{\cal N}(1,2)\ket{\kvec'\uparrow , - \kvec'\downarrow}_a\,.
\label{eq:Pdef}\end{eqnarray}
where ${\cal W}(1,2)$ and ${\cal N}(1,2)$ are non-local, energy
independent two-body operators. The dominant, local contributions are
in momentum space
\begin{equation}
  \tilde {\cal N} (k) = \tilde\Gamma_{\rm dd}(k)\,,\qquad
  \tilde {\cal W} (k) = -t(k)\tilde\Gamma_{\rm dd}(k)\SF^{-1}(k)
  \,.
  \label{eq:Veff}
  \end{equation}

With the result (\ref{eq:Ebcs}), we have arrived at a formulation of
the theory which is isomorphic to the BCS theory for weakly
interacting systems. Note that ${\cal W}(1,2)$ should be identified
with a static approximation of the $T$-matrix \cite{cbcs}.

\section{Results and Discussion}
\label{sec:results}

Before we discuss our results we should comment on the expected
accuracy. There are two important aspects: The first is the
convergence of the FHNC//n hierarchy as a function of density, and the
second is the dependence of the convergence on the interaction
strength. We have checked these issues in previous work for several
cases: In systems characterized by the Lennard-Jones
potential, we have found \cite{ljium} that the FHNC//0 approximation
for the energy is accurate within a percent below about 25 percent of
the saturation density of liquid \he3, $\rho_0\approx
0.274\,\sigma^{-3}$.  In the calculations to be reported below, we
went up to densities $\KF\sigma = 0.3$, corresponding to a density of
$\rho=0.0009\,\sigma^{-3}$.

A second and more difficult question is the convergence as a function of
interaction strength. The first bound state of the P\"oschl-Teller
potential appears at $V_0 = 2$ where the vacuum scattering length
diverges. We went in our calculations up to a value of $V_0 = 1.8$,
corresponding to a scattering length of $a_0 = -4.524\,\sigma$. This
is still small compared to the scattering length of $a_0\approx
-18\,$fm for neutron matter where we have compared the FHNC//0
approximation with the full FHNC scheme and found excellent agreement
\cite{ectpaper}.

All of this refers, of course, to the energy. It is well-known that
the energy is relatively insensitive to the accuracy of the
correlation functions and there is no guarantee that other quantities
come out just as accurately. Extensive investigations on these issues
are under way \cite{FanThesis} and will be published elsewhere.

\subsection{Energetics}
\label{ssec:energy}

A condition for the existence of solutions of the Euler equation is
that the term under the square-root in Eq. (\ref{eq:FermiRPA0})
is positive, or, in CBF, that the static density-density response
function is positive. This is expressed in terms of Landau's stability
criterion $F_0^s > -1$, and we must identify the limit
\begin{equation}
\tilde V_{\rm p-h}(0+) = \frac{\hbar^2\KF^2}{3m}F_0^s \equiv m(c^2-c_{\rm F}^2)
\label{eq:mcfromVph}
\end{equation}
where $c_{\rm F}^2 = \frac{\hbar^2\KF^2}{3m^2}$ is the speed of sound
of the free Fermi gas. The Landau-parameter $F_0^s$ can also be
obtained from the equation of state
\begin{equation}
mc^2 = \frac{d}{d\rho}\rho^2 \frac{d}{d\rho}\frac{E}{N}\,.
\label{eq:mcfromeos}
\end{equation}
The values of $mc^2$ obtained via Eq. (\ref{eq:mcfromeos}) and from a
diagrammatic expansion of the particle-hole interaction via
(\ref{eq:mcfromVph}) agree only in an exact theory \cite{parquet5},
their discrepancy can be used as a convergence test. We have checked
this by fitting the energy per particle by the kinetic energy
$3\hbar^2\KF^2/10m$ {\em plus\/} a polynomial of $\KF^3$ and $\KF^5$
and compared the results obtained from Eqs. (\ref{eq:mcfromeos}) and
(\ref{eq:mcfromVph}). We found that the numerical values are
practically identical for weak couplings. They can differ by about 30
percent for strong coupling strengths $V_0 > 1.7$, consistent with the
fact that the convergence of cluster expansions becomes worse with
increasing interaction strength.

\begin{figure}
  \centerline{\includegraphics[width=0.35\textwidth,angle=-90]{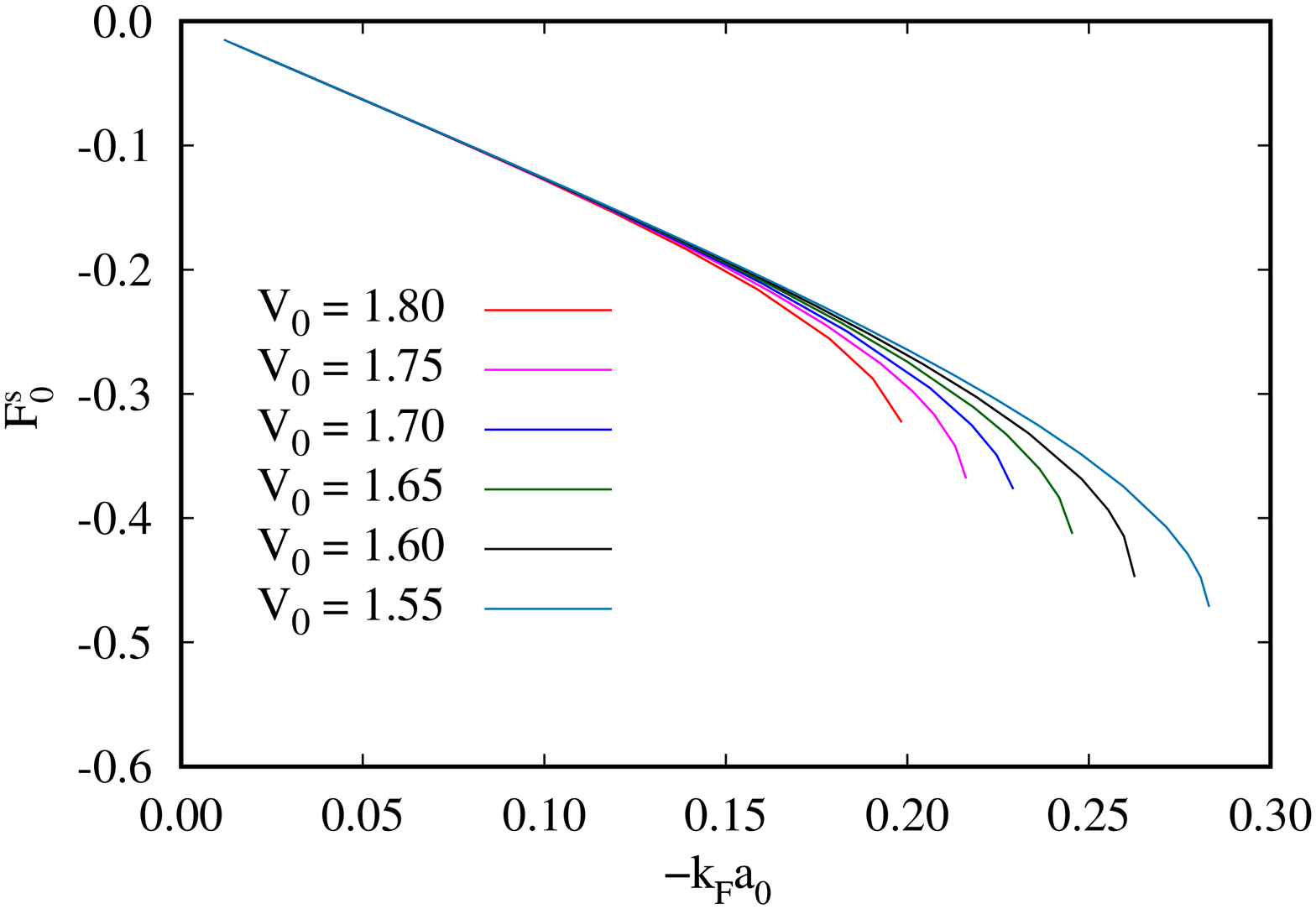}
  \includegraphics[width=0.35\textwidth,angle=-90]{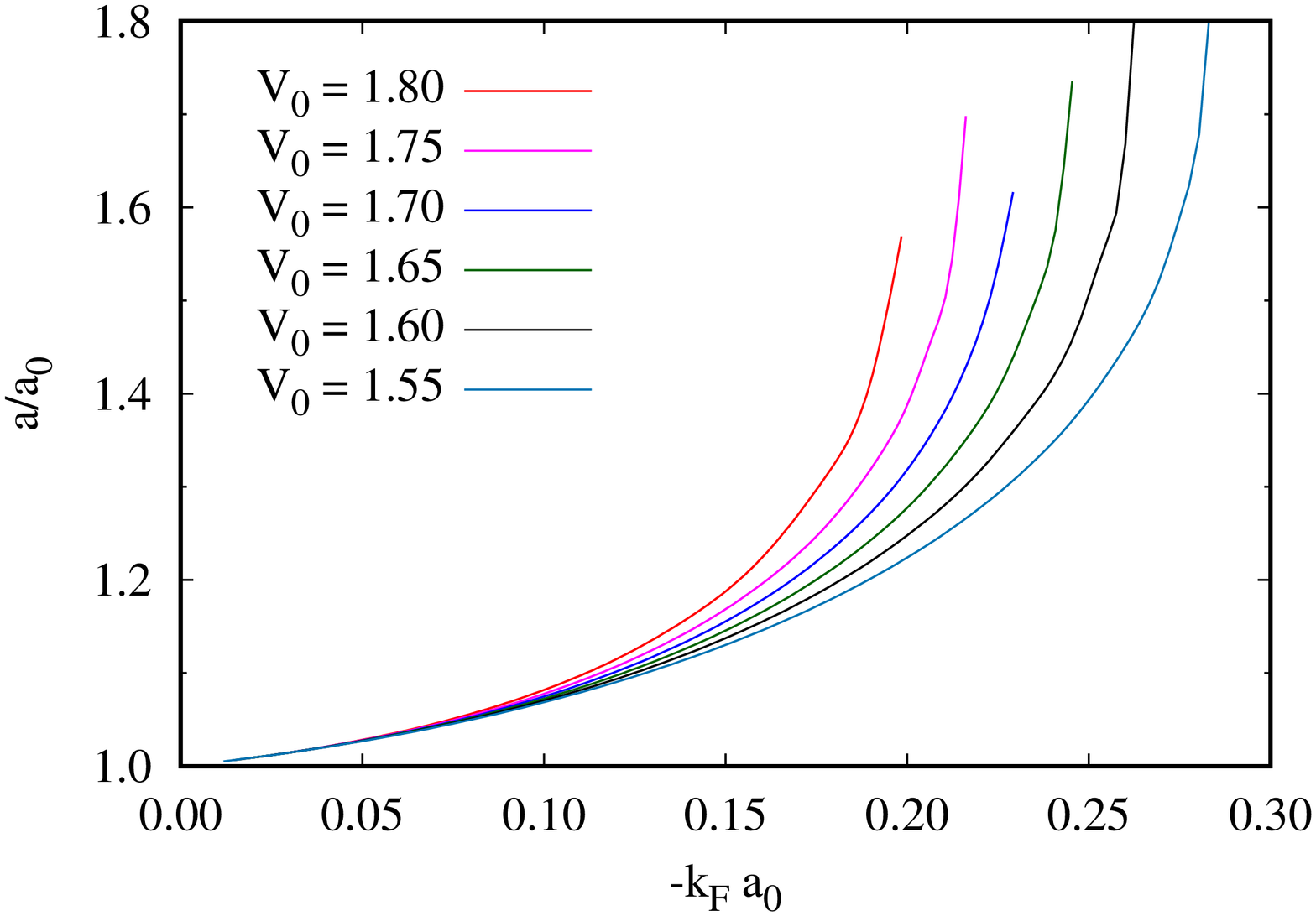}}
  \caption{(color online) The left figure shows the Fermi-liquid
    parameter $F_0^s$ of the ``P\"oschl-Teller'' gas for a sequence
    of interaction strengths $V_0$ as shown in the legend as a
    function of density, as calculated from Eq. (\ref{eq:mcfromVph}).
    The right figure shows the ratio between the in-medium scattering
    length $a$ and the vacuum scattering length $a_0$ for the same
    sequence of interaction strengths.\label{fig:eosplots}}
  \end{figure}

Let us now turn to the energetics and stability of the system. The
left panel of Fig. \ref{fig:eosplots} shows the Fermi-liquid parameter
$F_0^s$ as obtained from Eq. (\ref{eq:mcfromVph}) for a sequence of
interaction strengths as a function of the density.
The fact that the equations of state all come to an endpoint has been
identified in Ref. \citenum{cbcs} as due to a divergence of the
{\em in-medium scattering length.\/} We have mentioned already above that the
two--body operator ${\cal W}(1,2)$ should be identified with a static
approximation to the $T$-matrix which is, in the local approximation
used here, given by Eq. (\ref{eq:Veff}). Following the derivation of
Ref. \citenum{PethickSmith} of the low--density limit of the
superfluid gap in terms of the vacuum scattering length leading to
their Eq. (16.91), we find we can write the solution in exactly the same
form by replacing the vacuum scattering length $a_0$ by
\begin{equation}
a \equiv \frac{m}{4\pi\rho\hbar^2}\tilde {\cal W}(0+)\,.
\label{eq:amedium}
\end{equation}
which we therefore identify with the {\em in-medium\/} scattering length.
Note that, of course, $a\rightarrow a_0$ as $\rho\rightarrow 0$.

This divergence is the reason that the Landau stability limit $F_0^s
\rightarrow -1$ could not be reached. The same situation occurs,
expectedly, in the present case, see the right panel in
Fig. \ref{fig:eosplots}.  Due to this instability we have not been
able to reach the rather large values of $-\KF\a0$ reported in
Ref. \citenum{GC2008} before the optimization of the correlations
diverged.

\subsection{BCS pairing}
\label{ssec:pairing}

The wave function (\ref{eq:wavefunction}) determines the pairing
interaction uniquely. Since the effective interaction (\ref{eq:Veff})
contains chain diagrams, the important polarization effects
\cite{CKY76,SPR2001} are included in the density channel in a static
manner.  We have, however, pointed out in section \ref{ssec:CBF} that
this approximation can be improved by replacing the ``collective
approximation'' by the proper Lindhard function. Again, we take this
here as a plausibility argument; the rigorous derivation that such a
procedure is legitimate can be obtained by deriving the generalization
of the expansion (\ref{eq:Ebcs}) in correlated basis functions
\cite{CBFPairing}.

\begin{equation}
  \tilde V_{\rm eff}(k,\omega) = \tilde V_{\rm p-h}(k)
         \left[1-\chi_0(k,\omega)\tilde V_{\rm p-h}(k)\right]^{-1}
\end{equation}
which we take, following Ref. \citenum{SPR2001} at $\omega=0$.

\begin{figure}
  \centerline{\includegraphics[width=0.35\textwidth,angle=-90]{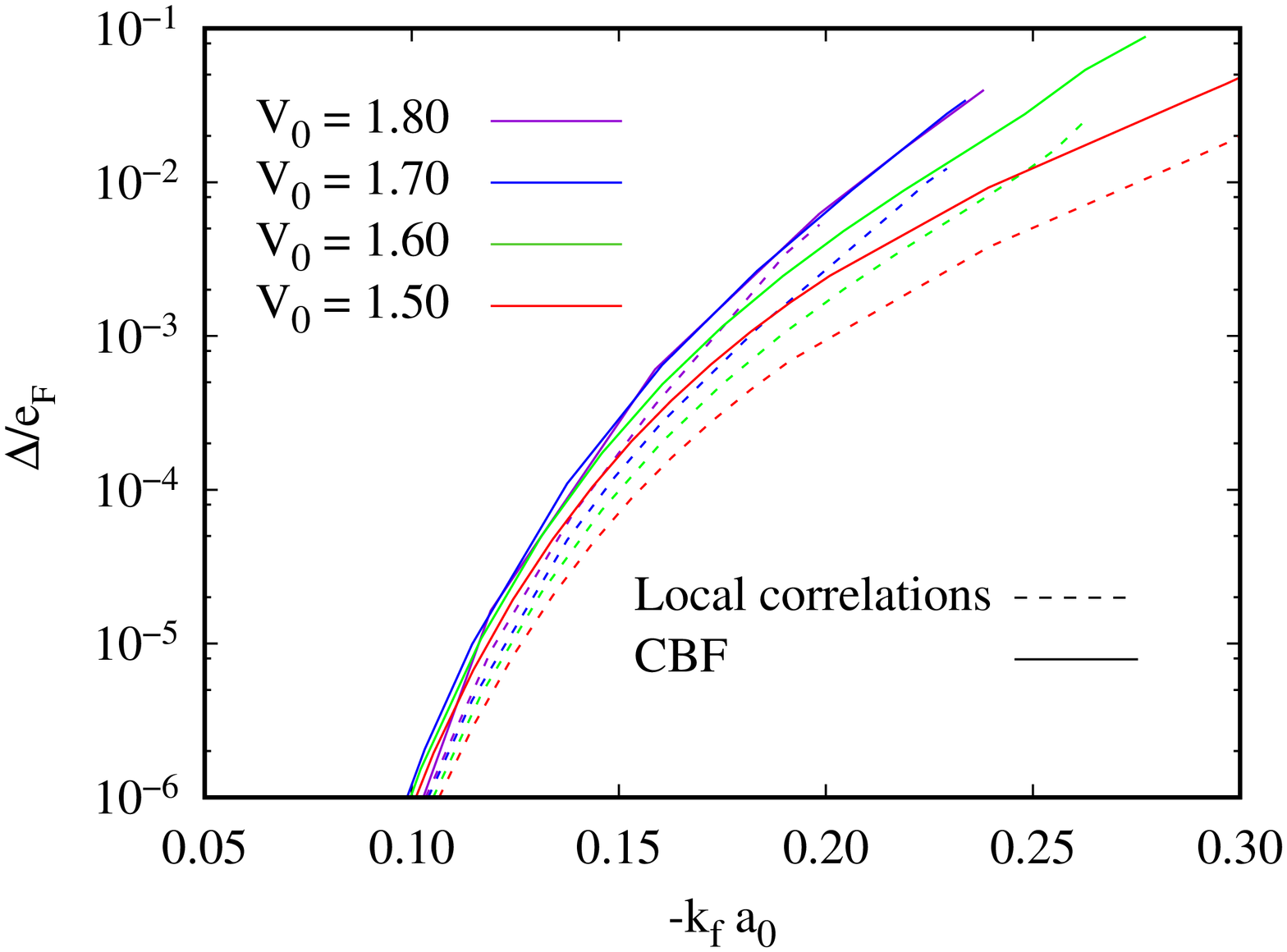}
  \includegraphics[width=0.35\textwidth,angle=-90]{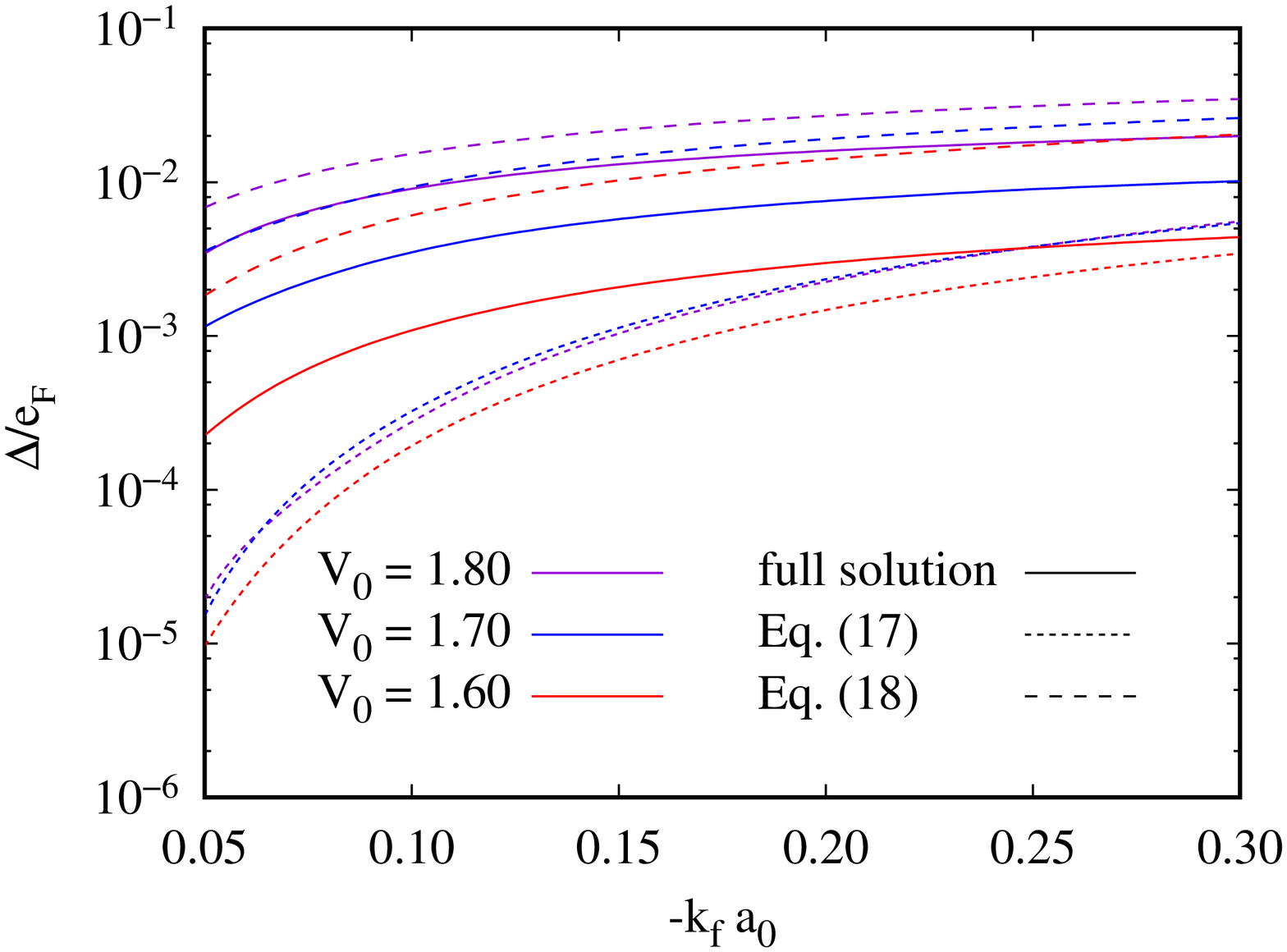}}
\caption{(color online) The left figure shows the superfluid gap, in
  units of the Fermi energy $e_F$ of the non-interacting system, for a
  number of coupling strengths in CBF (solid lines) and in FHNC//0-EL
  approximation (dashed lines). The CBF results for $V_0=1.70$ and
  $V_0 = 1.80$ are almost indistinguishable. The right figure shows
  the the gap as calculated from the bare P\"oschl-Teller potential
  (solid lines) as well as the one obtained in the low-density limit,
  Eq. (\ref{eq:spgap}) and the weakly coupled limit,
  Eq. (\ref{eq:fwgap}).
\label{fig:manygaps}
 }
\end{figure}

Fig.~\ref{fig:manygaps} shows the calculated energy gap in FHNC//0-EL
and CBF approximation. Evidently, inclusion of the energy-dependent
effective interaction can change the value of the gap by almost an
order of magnitude. This is, of course, not a statement on the
specific FHNC approximation, but more generally on the quality of the
locally correlated wave function which must, therefore, be seriously
questioned.

Since the P\"oschl-Teller potential does not have a hard core,
it is also possible to solve the gap equation without
correlations, and compare the result with different popular
approximate solutions. In the limit of low densities, the
gap at the Fermi surface is given by \cite{PethickSmith}
\begin{equation}
\frac{\Delta}{\EF} = \frac{8}{e^2}\exp\left(\frac{\pi}{ 2a_0\KF}\right)\,.
\label{eq:spgap}
\end{equation}
where $\EF = \hbar^2\KF^2/2m$, and $a_0$ is the vacuum scattering
length. At higher densities, but weak coupling, one has
\cite{FetterWalecka}
\begin{equation}
  \frac{\Delta}{\EF} \approx 8\exp\left(\frac{\pi\EF}
       {\bra{\KF}V\ket{\phi_{\KF}}}\right)\,,\qquad
  \bra{\KF}V\ket{\phi_{\KF}}
    = \KF\int_0^\infty dr \sin^2(r\KF) V(r)\,.
       \label{eq:fwgap}
\end{equation}
We show in the right panel of Fig. \ref{fig:manygaps} the gap as
obtained from the solution of the full equation as well as the two
approximations (\ref{eq:spgap}) and (\ref{eq:fwgap}).  At relatively
high densities, one obtains roughly the same order of magnitude,
whereas there are significant deviations at low densities. Evidently
one must go to much lower densities to recover the limit
(\ref{eq:spgap}); this was done in Ref. \citenum{cbcs}.

\subsection{Conclusion}

We have described in this paper new calculations of stability regime
and the pairing gap in a model system interacting via the attractive
P\"oschl-Teller interaction.  Similar to what we found in previous
work \cite{cbcs}, we have encountered an instability of the system
with increasing density and increasing potential strength $V_0$, well
before the vacuum scattering length $\a0$ of the interaction potential
diverges.

We have demonstrated that local correlation functions perform
poorly for pairing phenomena: The plausible reason for that is that
the wave function (\ref{eq:Jastrow}) treats all particles in the same
way.  This is a reasonable assumption for Fermi--sea averaged
quantities like the energy per particle or the static structure
function. However, this approximation is particularly poor for
observables that are determined by the dynamics close to the Fermi
surface. Since this is the case for BCS type pairing, our results are
as expected and fully consistent with our earlier work
\cite{shores,CCKS86}.

In conclusion, we note that going beyond the ``weak coupling''
approximation (\ref{eq:Ebcs}) makes very little difference in our
results but causes a number of serious formal difficulties
\cite{FanThesis}.  This is an interesting observation {\em per-se:\/}
Note that the gap equation, when treated at the level of a mean--field
theory, can describe the transition from a ``BCS'' state where the
Cooper pairs are weakly coupled to a ``BEC'' phase where the pairs are
strongly bound \cite{NozieresSchmittRink}. Solving a gap equation with
a self--consistently determined paring interaction ${\cal
  P}_{\kvec,\kvec'}$ is also a part of our diagram summation, but the
resulting Bogoliubov amplitudes then also define the superfluid
propagators used to sum the parquet diagrams which shows the same
instability as the summation of the same diagrams in the normal
system. The issue deserves further investigation.

\begin{acknowledgements}

This work was supported, in part, by the College of Arts and Sciences,
University at Buffalo SUNY, and the Austrian Science Fund project I602
(to EK). Discussions with Jordi Boronat, John W. Clark, and Robert
E. Zillich are gratefully acknowledged.

\end{acknowledgements}


\end{document}